\documentclass[conference]{IEEEtran}
\IEEEoverridecommandlockouts

\usepackage{cite}
\usepackage{amsmath,amssymb,amsfonts}
\usepackage{algorithmic}
\usepackage{graphicx}
\usepackage{textcomp}
\usepackage{xcolor}
\usepackage{nccmath}
\usepackage{tikz}

\begin{document}
\title{Analytical Nonlinear Distortion Characterization for Frequency-Selective Massive MIMO Channels}

\author{\IEEEauthorblockN{Murat Babek Salman$^{\dagger}$, Emil Björnson*, Gökhan Muzaffer Güvensen$^{\dagger}$,Tolga Çiloğlu$^{\dagger}$} 
\IEEEauthorblockA{{$^{\dagger}$Dept. of Electrical and Electronics Eng., METU, Ankara, Türkiye} \IEEEauthorblockA{*School of Electrical Engineering and Computer Science, KTH Royal Institute of Technology, Kista, Sweden}}{mbsalman@metu.edu.tr, emilbjo@kth.se, guvensen@metu.edu.tr, ciltolga@metu.edu.tr}
}% <-this % stops a space
\maketitle

\begin{abstract}
Nonlinear distortion stemming from low-cost power amplifiers may severely affect wireless communication performance through out-of-band (OOB) radiation and in-band distortion.
The distortion is correlated between different transmit antennas in an antenna array, which results in a beamforming gain at the receiver side that grows with the number of antennas. In this paper, we investigate how the strength of the distortion is affected by the frequency selectivity of the channel. A closed-form expression for the received distortion power is derived as a function of the number of multipath components (MPCs) and the delay spread, which highlight their impact. The performed analysis, which is verified via numerical simulations, reveals that as the number of MPCs increases, distortion exhibits distinct characteristics for in-band and OOB frequencies. It is shown that the received in-band and OOB distortion power is inversely proportional to the number of MPCs, and it is reported that as the delay spread gets narrower, the in-band distortion power is beamformed towards the intended user, which yields higher received in-band distortion compared to the OOB distortion. 
\end{abstract}


% \fi
\IEEEpeerreviewmaketitle

\section{Introduction}

Massive multiple-input multiple-output (MIMO) systems play a crucial role in increasing the spectral efficiency of modern communication systems \cite{7402270}. However, the cost-efficient employment of massive array structures restricts the quality of the hardware components equipped at the base station (BS). To reduce the hardware cost, reduced-quality components are preferred to be used in the BS \cite{6891254}. Consequently, low-quality hardware yields nonlinear distortion on the transmitted signal, which generates both out-of-band (OOB) and in-band distortion. The OOB radiation power is a measure of the additional interference power in adjacent frequency bands, which is limited by regulations and standards \cite{Standards}. In-band distortion, on the other hand, deteriorate the signal quality at the intended receiver, which reduces the detection performance.

The nonlinear distortion radiated from an antenna array is studied in  \cite{8186238,9113723, 7303917,Mollen1,Mollen2,8917597}. In \cite{8186238} and \cite{9113723}, the impact of the distortion on the received signal quality are investigated. Analytical bit-error-rate (BER) expressions are derived in \cite{7303917,8186238,9113723} to quantify the effect of in-band distortion. In these studies, the nonlinear distortion is accurately characterized for per antenna, but approximated as uncorrelated between the antennas, which makes it radiated isotropically. However, as the input signals of each transmit antenna is correlated when performing beamforming, there is correlation also between the distortion signals and is particularly visible in massive MIMO. It is experimentally shown in \cite{7414011} that the nonlinear distortion is also (partially) coherently beamformed towards the users due to its correlated nature. Therefore, per-antenna distortion analysis is not sufficient to fully capture the spatial distortion characteristics. 

The spatial distribution of OOB radiation is investigated by including the distortion correlation in \cite{Mollen1}, where it is shown numerically that the distortion power is also beamformed towards the intended user; however, its beamforming gain is smaller than the desired signal. In \cite{Mollen2}, a mathematical analysis on OOB radiation pattern is performed to analyze the spatial distribution by using It\^{o}-Hermite polynomials. Multi-user and frequency selective fading scenarios are investigated through a semi-analytical approach. It is shown that for a single user with a frequency flat channel, the distortion becomes almost fully correlated, and the distortion power is coherently beamformed to the intended user. However, as the number of users or multipath components (MPCs) increases, distortion spreads over the spatial domain; hence, the received distortion power gets weakened. By using the analytical framework from \cite{Mollen2}, a semi-analytical BER expression is derived in \cite{8917597} by using the in-band distortion statistics.

This paper analyzes the effects of frequency-selective channels on the OOB and in-band distortion characteristics in a single-user scenario. By exploiting the assumption of a rich scattering environment and optimal maximum ratio transmission (MRT), we derive a closed-form expression for the OOB and in-band distortion received at the user. The analytical distortion power expression includes channel characteristics such as the number of MPCs, the delay spread of the channel, and the number of transmit antennas. To the authors' knowledge, this paper provides the first closed-form expression for the received distortion power, which includes the MPC distribution as a variable. It is discovered that depending on the number of MPCs and the delay spread, the distortion varies differently for in-band and OOB frequencies. This kind of distortion behavior is not reported in previous literature. The derived expressions are verified via Monte-Carlo simulations by showing that the analytical power spectral density (PSD) of the received signal and error vector magnitude estimates are well in line with the numerical results. Estimated received distortion power expression is useful for determining the system parameters and evaluating the system performance.

\section{System Model}

This paper considers a cellular system where a BS with $M$ antennas is serving a single user in a rich scattering environment. A frequency-selective block fading channel is assumed such that channel realization within each block is stationary. In this paper, we denote signals in the discrete domain; however, to approximate the continuous-time operations, we consider oversampled signals with oversampling factor $\mu$. According to these assumptions, the transmitted orthogonal frequency-division multiplexing (OFDM) modulated signal can be expressed as in time domain as

\begin{equation}
    {\bf x}[n] = \sum_{k=-N_s/2+1}^{N_s/2} {\bf w}_k a_k e^{j\frac{2 \pi}{\mu N_s} k n},
\end{equation}
where ${\bf w}_k \in \mathbb{C}^{M \times 1} $ is the precoding vector at the $k^{\rm th}$ subcarrier, $\{a_k\}$'s are the quadrature amplitude modulated (QAM) digital information signals, which are non-zero for $k = -N_s/2+1,\ldots, N_s/2$, and $N=\mu N_s$ is the symbol duration. Before radiating through the air, the modulated signal is amplified via PAs to increase the signal power to a sufficient level. However, the nonideality of the PA hardware causes nonlinear distortion. For the sake of generality, we represent the nonlinearity for $m^{\rm th}$ antenna element as ${\Tilde{x}}^{(m)}[n]=\Psi(x^{(m)}[n] )$.
%\begin{equation}
 %   {\Tilde{x}}^{(m)}[n] = \sum_{p=1,p:\textrm{odd}}^{P}  \omega_{pm} x^{(m)}[n] |x^{(m)}[n]|^{(p-1)},
%\end{equation}
%
%where $\omega_{pm}$'s are the polynomial basis coefficients. 
The transmitted signal from the antenna array becomes ${\Tilde{\bf x}}[n] = [{\Tilde{x}}^{(1)}[n] , \ldots, {\Tilde{x}}^{(M)}[n]]^T$. Then the received signal at the single-antenna user terminal can be expressed as

\begin{equation}
    y[n] = \sum_{l=0}^{L-1} {\bf h}^T[\tau_l] {\Tilde{\bf x}}[n-\tau_l],
\end{equation}
where ${\bf h}[l] \in \mathbb{C}^{M \times 1}$ is the Rayleigh fading communication channel whose $m^{\rm th}$ element is ${ h}^{(m)}[l] \sim \mathcal{N}_{\mathbb{C}}(0,1)$, and $L$ is the number of significant taps. The channel vector can be equivalently represented in the frequency domain as

\begin{equation}
    {\bf h}_k = \sum_{l=0}^{L-1} {\bf h}[\tau_l] e^{-j\frac{2 \pi}{\mu N_s} k \tau_l},
\end{equation}
for $k = -N/2+1,\ldots, N/2$. Then we can define the precoding vector for MRT as ${\bf w}_k = {\bf h}^*_k/\sqrt{L}$, which is optimal for single-user transmission. %Note that the other types of precoding vectors can also be adopted; however, in this study, MRT precoding is preferred since it provides analytical tractability. 

\section{Distortion Characterization under Frequency-Selective Channels}

\subsection{Spectral characterization of the nonlinear distortion}

It is known that if the number of subcarriers is high enough, the distribution of OFDM modulated signals can be approximated by the circularly symmetric complex Gaussian distribution thanks to the central limit theorem. By exploiting this fact, we can adopt a complex Itô-Hermite polynomial representation to acquire the second-order statistics of the nonlinear distortion in the frequency domain \cite{Mollen2}. Consequently, the nonlinearly transmitted signal is rewritten by focusing on only the third-order nonlinearity as

\begin{equation}
    {\Tilde{x}}^{(m)}[n] = \underbrace{\alpha_{1m} x^{(m)}[n]}_{\textrm{Desired signal : }u^{(m)}[n] } + \underbrace{\alpha_{3m} \sigma_{x^{(m)}}^3 H_3 \left( \frac{x^{(m)}[n]}{\sigma_{x^{(m)}}} \right)}_{\textrm{Distortion signal : } d^{(m)}[n] }, \label{HermiteExpansion}
\end{equation}
where $H_3(\cdot)$ is the third-order Hermite basis function

\begin{equation}
    H_3(x) \triangleq \sum_{i=0}^{1} (-1)^i i! \begin{pmatrix} 2 \\ i \end{pmatrix} \begin{pmatrix} 1 \\ i \end{pmatrix} x |x|^{2-2i},
\end{equation}
and $\sigma_{x^{(m)}}$ is the standard deviation of $x^{(m)}$.

This representation has the following useful property, which enables spectral analysis. The desired and the distortion signals are uncorrelated $\mathbb{E}\left[ u^{(m)}[n] (d^{(m')}[n-n'])^* \right]=0$ $\forall n',m,m'$, so that we can write the cross-spectral density as the summation of two uncorrelated signal terms as

\begin{equation}
    S_{{\tilde{x}^{(m)}} {\tilde{x}^{(m')}}}[k] = \alpha_{1m} \alpha_{1m'}^* S_{x^{(m)}x^{(m')}}[k] + S_{d^{(m)} d^{(m')}}[k], \label{Stilde}
\end{equation}
where $S_{x^{(m)}x^{(m')}}[k] = \mathcal{F}\left \{ R_{x^{(m)} x^{(m')}}[n-n'] \right\}$ is the cross-spectral density for the desired term, which is defined by the Fourier transform of cross-correlation function $R_{x^{(m)} x^{(m')}}[n-n'] = \mathbb{E} \left[x^{(m)}[n] x^{(m')}[n'] \right]$. Similarly, $S_{d^{(m)} d^{(m')}}[k]$ is the Fourier transform of the cross-correlation function of the distortion term, which can be expressed by exploiting the Gaussianity of the PA input signals as

\begin{equation}
\begin{split}
    &R_{d^{(m)} d^{(m')}}[n-n']\\
    & = 2\alpha_{3m} \alpha_{3m'}^* R_{x^{(m)}x^{(m')}}[n-n'] \left|R_{x^{(m)}x^{(m')}}[n-n'] \right|^2. \label{DistCorr}
\end{split}
\end{equation}

By using the expression in \eqref{DistCorr}, cross-spectral density $S_{d^{(m)} d^{(m')}}[k]$ can be obtained as

\begin{equation}
\begin{split}
    &S_{d^{(m)} d^{(m')}}[k]=  2\alpha_{3m}\alpha_{3m'}^*\\
    & \qquad  \left( S_{x^{(m)}x^{(m')}}[k] \ast S_{x^{(m)}x^{(m')}}[k] \ast S_{x^{(m)}x^{(m')}}^*[-k] \right) \label{Sdd},
\end{split}
\end{equation}
where $\ast$ denotes the convolution operation. Besides, $S_{x^{(m)}x^{(m')}}[k]$ can be written by using the precoding vector for a given channel as

\begin{equation}
    S_{x^{(m)}x^{(m')}}[k] = \frac{\left(h_k^{(m)}\right)^*  \left(h_k^{(m')}\right)}{L}, \quad |k|\leq N_s/2. \label{Sxx}
\end{equation}
By using \eqref{Sdd} and \eqref{Sxx}, we can express the cross-spectrum of the transmitted distortion signal as

\begin{equation}
\begin{split}
    &S_{d^{(m)} d^{(m')}}[k]= \frac{2}{N_s^2} \frac{\alpha_{3m} \alpha_{3m'}^*}{L^3}\sum_{k' = -N/2+1}^{N/2} \sum_{k'' = -N/2+1}^{N/2}\\
    &\Bigg[  \left(h_{k'}^{(m)}\right)^* \left(h_{k''}^{(m)}\right)^* h_{k'+k''-k}^{(m)}   h_{k'}^{(m')} h_{k''}^{(m')} \left(h_{k'+k''-k}^{(m')}\right)^*\ \Bigg], \label{SddTx}
\end{split}
\end{equation}
for $|k'|\le N_S/2$, $|k''|\le N_S/2$ and $ |k'+k''-k|\le N_S/2$. 
\subsection{Spectral analysis of the received distortion}

In this section, we will derive the spectral characteristics of the received distortion signal. The received power by the intended user at the subcarrier $k$ is

\begin{equation}
   S_{yy}[k] = \sum_{m=1}^{M} \sum_{m'=1}^{M} h_k^{(m)} S_{{\tilde{x}^{(m)}} {\tilde{x}^{(m')}}}[k] \left(h_k^{(m')}\right)^*.
\end{equation}

By utilizing the decomposition of $S_{{\tilde{x}^{(m)}} {\tilde{x}^{(m')}}}[k]$ in \eqref{Stilde}, we can rewrite the PSD of the received signal as

\begin{equation}
    S_{yy}[k] = S_{uu}[k] + S_{dd}[k],
\end{equation}
where $S_{uu}[k]$ is the PSD of the desired term, which is expressed as

\begin{equation}
\begin{split}
    S_{uu}[k] = \left|S_{u}[k]\right|^2
    %&\underbrace{\left( \sum_{m=1}^{M} \frac{\alpha_{1m}}{\sqrt{L}} h_{k}^{(m)} \left(h_{k}^{(m)} \right)^* \right)}_{S_{u}[k]}\\ &\underbrace{\left( \sum_{m'=1}^{M} \frac{\alpha_{1m'}}{\sqrt{L}} h_{k}^{(m')} \left(h_{k}^{(m')} \right)^* \right)^*}_{S_{u}^*[k]},
    \end{split}
\end{equation}
where $S_{u}[k]=\left( \sum_{m=1}^{M} \frac{\alpha_{1m}}{\sqrt{L}} h_{k}^{(m)} \left(h_{k}^{(m)} \right)^* \right)$ and $S_{dd}[k]$ is the PSD of the uncorrelated distortion term, which is obtained as
%\begin{equation}
  %  \begin{split}
      %  &S_{dd}[k]= \frac{2}{N_s^2} \sum_{k' = -N/2+1}^{N/2} \sum_{k'' = -N/2+1}^{N/2}\\
      %  &\underbrace{\left( \sum_{m=1}^{M}\alpha_{3m} \frac{1}{L\sqrt{L}} h_{k}^{(m)} (h_{k'}^{(m)})^* (h_{k''}^{(m)})^* h_{k'+k''-k}^{(m)} \right)}_{S_{d}[k,k',k'']} \\
      %  &\underbrace{\left(\sum_{m'=1}^{M} \alpha_{3m'} \frac{1}{L\sqrt{L}} h_{k}^{(m')} \left(h_{k'}^{(m')}\right)^* \left(h_{k''}^{(m')}\right)^* h_{k'+k''-k}^{(m')} \right)^*}_{S_{d}^*[k,k',k'']}, 
    %\end{split}
%\end{equation}
\begin{equation}
    S_{dd}[k]= \frac{2}{N_s^2} \sum_{k' = -N/2+1}^{N/2} \sum_{k'' = -N/2+1}^{N/2} \left|S_{d}[k,k',k''] \right|^2
\end{equation}
where 
\begin{equation}
    S_{d}[k,k',k'']= \sum_{m=1}^{M} \frac{\alpha_{3m}}{L\sqrt{L}} h_{k}^{(m)} (h_{k'}^{(m)})^* (h_{k''}^{(m)})^* h_{k'+k''-k}^{(m)}
\end{equation}
for $|k'|\le N_S/2$, $|k''|\le N_S/2$ and $ |k'+k''-k|\le N_S/2$. Note that the received PSD expression depends on the instantaneous channel realizations. However, we can exploit the massive spatial diversity achieved when having a large number of antennas to obtain a deterministic closed-form expression for the received signal and distortion power.

As the first step, we will obtain the expression for the power of the desired signal term. Consider the term

\begin{equation}
    S_{u}[k] =  \sum_{m=1}^{M} \frac{\alpha_{1m}}{\sqrt{L}} h_{k}^{(m)} \left(h_{k}^{(m)} \right)^*,
\end{equation}
which is the summation of many independent random variables; hence, the law of large numbers (LLN) implies that $\frac{S_{u}[k]}{M}\rightarrow \mathbb{E} [ \alpha_{1m}  | h_k^{(m)} |^2 / \sqrt{L} ]$ when $M \to \infty$. For a large but finite number of antennas, we can use the LLN to approximate the random variable $S_{u}[k]$ by a single deterministic scalar as
\begin{equation}
    S_{u}[k] \approx M \mathbb{E} \left[ \alpha_{1m} {\left| h_k^{(m)} \right|^2}/{\sqrt{L}}\right].
\end{equation}
By using the uncorrelated Rayleigh fading channel assumption, we can simplify this expression as
\begin{equation}
    S_{u}[k] \approx \alpha_{1} M \sqrt{L}, \label{Sut}
\end{equation}
where $ \alpha_{1} =  \mathbb{E} \left[ \alpha_{1m} \right]$ and $\mathbb{E} [ | h_k^{(m)} |^2] = L $ by using Parseval's theorem. By using \eqref{Sut}, we can find the power of the desired term in the received signal $S_{uu}[k]$ as

\begin{equation}
    S_{uu}[k] \approx |\alpha_1|^2M^2L. \label{Pu}
\end{equation}

Having obtained the power of the desired signal, we can focus on the power of the distortion term. In order to follow the same approach as in the desired term, consider the expression

\begin{equation}
   S_{d} [k,k',k''] =  \sum_{m=1}^{M}\underbrace{ \frac{\alpha_{3m}}{L\sqrt{L}} h_{k}^{(m)} \left(h_{k'}^{(m)}\right)^* \left(h_{k''}^{(m)}\right)^* h_{k'+k''-k}^{(m)}}_{S_{d}^{(m)} [k,k',k'']},
\end{equation}
which can be similarly approximated by using the LLN property $\frac{S_{d} [k,k',k'']}{M}\rightarrow \mathbb{E} \left[S_{d}^{(m)} [k,k',k''] \right]$  as

\begin{equation}
    S_{d}[k,k',k''] \approx  \frac{M \alpha_3}{L\sqrt{L}} \mathbb{E} \left[ h_{k}^{(m)} \left(h_{k'}^{(m)}\right)^* \left(h_{k''}^{(m)}\right)^* h_{k'+k''-k}^{(m)} \right]. \label{Sd}
\end{equation}

Calculation of this expectation is not straightforward and depends on the channel's frequency selectivity. To find the correlation between the subcarriers' channels in \eqref{Sd}, consider the equivalent representation of the expectation as

\begin{equation}
\begin{split}
    &\mathcal{E}_{k,k',k''}=\mathbb{E} \left[ h_{k}^{(m)} \left(h_{k'}^{(m)}\right)^* \left(h_{k''}^{(m)}\right)^* h_{k'+k''-k}^{(m)} \right]\\
    &= \sum_{\substack{\tau_l,\tau_{l'},\\\hat{\tau}_l,\hat{\tau}_{l'}} \in \mathcal{L}} \mathbb{E} \left[  h^{(m)}[\tau_{l}] \left(h^{(m)}[\tau_{l'}]\right)^* \left(h^{(m)} [\hat{\tau}_{l}]\right)^* h^{(m)}[\hat{\tau}_{l'}]  \right]\\
    &\qquad \qquad \qquad \cdot e^{-j\frac{2\pi}{N}k\tau_{l}} e^{j\frac{2\pi}{N}k'\tau_{l'}} e^{j\frac{2\pi}{N}k'' \hat{\tau}_{l}} e^{-j\frac{2\pi}{N}(k'+k''-k) \hat{\tau}_{l'}}, \label{Expect}
\end{split}
\end{equation}
where $\mathcal{L} \subset \mathbb{Z}^{L}$ is the set of MPC indices, $[\mathcal{L}]_l\in[0,\tau_{\max}] $. Note that the different multipath taps are uncorrelated; therefore, only a subset of the terms in \eqref{Expect} are non-zero, which can be expressed as in \eqref{Expect1} on the top of the next page.
\begin{figure*}
\begin{equation}
\begin{split}
    \mathcal{E}_{k,k',k''}=&\sum_{\tau_{l} \in \mathcal{L}}  \mathbb{E} \left[  \left| h^{(m)}[\tau_{l}] \right|^4 \right] +  \sum_{ \tau_{l} \in \mathcal{L} } \sum_{ \substack{\tau_{l'} \in \mathcal{L} \\ \ l\neq l'}} \left(\mathbb{E} \left[  \left| h^{(m)}[\tau_{l}] \right|^2 \right] e^{-j \frac{2\pi}{N}(k-k')\tau_l}  \right) \left( \mathbb{E} \left[  \left| h^{(m)}[\tau_{l'}] \right|^2 \right] e^{-j \frac{2\pi}{N}(k'-k)\tau_{l'}}\right)\\
    &+\sum_{ \tau_{l}} \sum_{ \substack{\tau_{l'} \in \mathcal{L} \\ \ l\neq l'}}  \left( \mathbb{E} \left[  \left| h^{(m)}[\tau_{l}] \right|^2 \right] e^{-j \frac{2\pi}{N}(k-k'')\tau_l} \right) \left( \mathbb{E} \left[  \left| h^{(m)}[\tau_{l'}] \right|^2 \right] e^{-j \frac{2\pi}{N}(k''-k)\tau_{l'}}\right) \label{Expect1}
\end{split}
\end{equation}
\hrulefill
\end{figure*}

From \eqref{Expect1}, it can be seen that the expectation depends on the difference between the subcarriers $k - k'$. Therefore, one can interpret $\mathcal{E}_{k,k',k''}$ as a correlation between the subcarriers. Furthermore, the expression of $\mathcal{E}_{k,k',k''}$ can be simplified as

\begin{equation}
\begin{split}
    &\mathcal{E}_{k,k',k''}\\
    &= \underbrace{\sum_{\tau_{l} \in \mathcal{L}} e^{-j \frac{2\pi}{N}(k-k')\tau_{l}}  \sum_{\tau_{l'} \in \mathcal{L}} e^{-j \frac{2\pi}{N}(k'-k)\tau_{l'}}}_{L^2f[k-k',\mathcal{L}]}\\
    & \quad + \underbrace{\sum_{\tau_{l} \in \mathcal{L}} e^{-j \frac{2\pi}{N}(k-k'')\tau_{l}}  \sum_{\tau_{l'} \in \mathcal{L}} e^{-j \frac{2\pi}{N}(k''-k)\tau_{l'}}}_{L^2f[k-k'',\mathcal{L}]}, \label{Expect2} 
    \end{split}
\end{equation}
which yields

\begin{equation}
    \mathcal{E}_{k,k',k''} = L^2f[k-k',\mathcal{L}] + L^2f[k-k'',\mathcal{L}]. \label{ExpFinal}
\end{equation}
Note that $f[k-k',\mathcal{L}]$ is a positive real number since it is the result of multiplication of a conjugate pair. It can be seen that $f[k-k',\mathcal{L}]$ is a function of the temporal distribution of multipaths, which causes  $f[k-k',\mathcal{L}]$ to be a random variable. However, as can be observed from \eqref{Pu}, the power of the desired term $S_{uu}[k]$ is independent of the MPC distribution. By using \eqref{ExpFinal}, the distortion power for a given MPC realization can be calculated as

\begin{equation}
    S_{dd}[k,\mathcal{L}] \approx  \frac{2M^2|\alpha_3|^2}{N_s^2L^3} \sum_{k',k''} (L^2f[k-k',\mathcal{L}] + L^2f[k-k'',\mathcal{L}])^2,
\end{equation}
for $|k'|\le N_S/2$, $|k''|\le N_S/2$ and $ |k'+k''-k|\le N_S/2$.

As mentioned above, we can treat the MPC distribution as a random variable; hence, 
We can find the average distortion power by taking the expectation with respect to the MPC distribution as

\begin{equation}
\begin{split}
S_{dd}[k] &\approx \mathbb{E}_{\mathcal{L}} \left[ S_{dd}[k,\mathcal{L}] \right]\\
  &\approx 2M^2\frac{|\alpha_3|^2}{N_s^2L^3} \sum_{k',k''}  \epsilon_{k,k'}^{f2} + 2\epsilon_{k,k',k''}^{ff} + \epsilon_{k,k''}^{f2} \label{AnalyticalDist}
   \end{split}
\end{equation}
%
%\begin{equation}
%\begin{split}
 %   &S_{dd}[k] \approx \mathbb{E}_{\mathcal{L}} \left[ S_{dd}[k,\mathcal{L}] \right]\\
  %  &= 2M^2 \frac{|\alpha_3|^2}{L^3} \sum_{k',k''} \underbrace{L^4 \mathbb{E}_{\mathcal{L}} \left[ f^2[k-k',\mathcal{L}] \right]}_{\epsilon_{k,k'}^{f2}}\\
  %  & \quad + 2\underbrace{L^4 \mathbb{E}_{\mathcal{L}} \left[  f[k-k',\mathcal{L}]   f[k-k'',\mathcal{L}] \right]}_{\epsilon_{k,k',k''}^{ff}}\ \\
  %  & \quad + \underbrace{ L^4 \mathbb{E}_{\mathcal{L}} \left[ f^2[k-k'',\mathcal{L}] \right]}_{\epsilon_{k,k''}^{f2}},
%\end{split}
%\end{equation}
%
for $|k'|\le N_S/2$, $|k''|\le N_S/2$ and $ |k'+k''-k|\le N_S/2$, where $\epsilon_{k,k'}^{f2} = L^4 \mathbb{E}_{\mathcal{L}} \left[ f^2[k-k',\mathcal{L}] \right]$, and $\epsilon_{k,k',k''}^{ff}= L^4 \mathbb{E}_{\mathcal{L}} \left[  f[k-k',\mathcal{L}]   f[k-k'',\mathcal{L}] \right]$. We can evaluate $\epsilon_{k,k',k''}^{ff}$ analytically assuming that the time delays $\tau_l$ have a particular random distribution. 
We obtain the expression \eqref{crossBins1}, on the top of the next page, when assuming that the time delays
are uniformly distributed independent random variables $\mathcal{U} [0,\tau_{\max}]$, where $\tau_{\max}$ is the maximum delay spread.

\begin{figure*}
    \begin{equation}
\begin{split}
    &\epsilon_{k,k',k''}^{ff} =  \sum_{l=0}^{L-1} \sum_{l'=0}^{L-1}  \sum_{{\hat l}=0}^{L-1} \sum_{{\hat l}'=0}^{L-1} \mathbb{E}_{\mathcal{L}}  \left[e^{-j\frac{2\pi}{N}(k-k')\tau_{l}}  e^{-j\frac{2\pi}{N}(k'-k)\tau_{l'}}  e^{-j\frac{2\pi}{N}(k-k'')\tau_{{\hat l}}}  e^{-j\frac{2\pi}{N}(k''-k)\tau_{{\hat l}'}} \right]
\end{split} \label{crossBins1}
\end{equation}
\end{figure*}

Note that for $l=l'={\hat l}={\hat l}'$, the term in the expectation in \eqref{crossBins1} becomes $1$; hence, we can rewrite this expectation by only focusing on the dominant terms, which results  in \eqref{Expect3}
\begin{figure*}
\begin{equation}
\begin{split}
    \epsilon_{k,k',k''}^{ff} \approx L^2 + \sum_{\substack{l\neq l' \neq {\hat l} \neq {\hat l}'}}^{L-1} \mathbb{E}_{\mathcal{L}} \left[ e^{-j\frac{2\pi}{N}(k-k')\tau_l} \right] \mathbb{E}_{\mathcal{L}} \left[ e^{-j\frac{2\pi}{N}(k'-k)\tau_{l'}} \right] \mathbb{E}_{\mathcal{L}} \left[ e^{-j\frac{2\pi}{N}(k-k'')\tau_{\hat l}} \right] \mathbb{E}_{\mathcal{L}} \left[ e^{-j\frac{2\pi}{N}(k''-k)\tau_{{\hat l}'}} \right], \label{Expect3}
    \end{split}
\end{equation}

\hrulefill
\end{figure*}
We also note that $\mathbb{E}_{\mathcal{L}} \left[ e^{-j\frac{2\pi}{N}(k-k')\tau_l} \right] \mathbb{E}_{\mathcal{L}} \left[ e^{-j\frac{2\pi}{N}(k'-k)\tau_{l'}} \right]$ is the multiplication of two complex conjugate numbers, so we can simplify  \eqref{Expect3} as

\begin{equation}
\begin{split}
        &\epsilon_{k,k',k''}^{ff}   \\
        &\approx  L^2 + \sum_{\substack{l,l',\\ {\hat l},{\hat l}'}}^{L-1} \bigg|\underbrace{ \mathbb{E}_{\mathcal{L}} \left[ e^{-j\frac{2\pi}{N}(k-k')\tau} \right]}_{\xi_{k-k'}} \bigg|^2  \bigg|\underbrace{\mathbb{E}_{\mathcal{L}} \left[ e^{-j\frac{2\pi}{N}(k-k'')\tau} \right]}_{\xi_{k-k''}}  \bigg|^2,\\
    & \approx L^2 +  (L^4-L^2) \left| \xi_{k-k'} \right|^2  \left| \xi_{k-k''}\right|^2.
\end{split}
\end{equation}
 Then the expectation for $\xi_{k-k'} $ can be evaluated as
 
 \begin{equation}
 \begin{split}
     \mathbb{E}_{\mathcal{L}}& \left[ e^{-j\frac{2\pi}{N}(k-k')\tau_l} \right] = \sum_{\tau=0}^{\tau_{\max}-1} \frac{1}{\tau_{\max}} e^{-j\frac{2\pi}{N}(k-k')\tau}\\
     &= \frac{1}{\tau_{\max}} \frac{e^{-j\frac{2\pi}{N}(k-k')\frac{\tau_{\max}}{2}}}{e^{-j\frac{2\pi}{N}(k-k')\frac{1}{2}}} \frac{ \sin\left( \frac{2\pi}{N} (k-k')  \frac{\tau_{\max}}{2}\ \right)  }{ \sin\left( \frac{2\pi}{N} (k-k')  \frac{1}{2}\ \right)}.
 \end{split}
 \end{equation}
Consequently $ \epsilon_{k,k'}^f \triangleq \mathbb{E}_{\mathcal{L}} \left[ L^2 f[k-k',\mathcal{L}] \right]$ becomes

\begin{equation}
     \epsilon_{k,k'}^f = L + \frac{L^2-L}{\tau_{\max}^2} \frac{\sin^2\left( \frac{2\pi}{N} (k-k')  \frac{\tau_{\max}}{2}\ \right)  }{\sin^2\left( \frac{2\pi}{N} (k-k')  \frac{1}{2}\ \right)}. \label{sinc1}
\end{equation}
Similarly, we can approximate $ \epsilon_{k'}^{f2} = L^4 \mathbb{E}_{\mathcal{L}} \left[ f^2[k-k',\mathcal{L}] \right]$ as

\begin{equation}
    \epsilon_{k,k'}^{f2} \approx L^2 + \frac{L^4-L^2}{\tau_{\max}^4} \frac{\sin^4\left( \frac{2\pi}{N} (k-k')  \frac{\tau_{\max}}{2}\ \right)  }{\sin^4\left( \frac{2\pi}{N} (k-k')  \frac{1}{2}\ \right)}. \label{sinc2}
\end{equation}

%Consequently, the distortion power for each frequency bin $k$ $S_{dd}[k]$ can be approximated as

%
%for $|k'|\le N_S/2$, $|k''|\le N_S/2$ and $ |k'+k''-k|\le N_S/2$.

In addition, by using the uncorrelatedness of the desired and the distortion signal terms, signal-to-distortion power ratio (SDR) for each data subcarrier $k$, $|k|<N_S/2$ can be given as

\begin{equation}
    {\rm{SDR}}[k] \approx \frac{|\alpha_1|^2M^2L}{2M^2 \frac{|\alpha_3|^2}{N_s^2L^3} \sum_{k',k''}  \epsilon_{k,k'}^{f2} + 2\epsilon_{k,k',k''}^{ff} + \epsilon_{k,k''}^{f2}}. \label{SDRExp}
\end{equation}

It is observed from \eqref{SDRExp} that both the numerator and the first term of the denominator are proportional with $M^2$. Consequently, the received distortion does not vanish as the number of transmit antennas increases, and it can be stated that the SDR becomes independent of $M$ when the number of antennas is large. However, as can be interpreted from \eqref{SDRExp}, the distortion power depends on the frequency selectivity via $\epsilon_{k,k'}^{f2}$ and $\epsilon_{k,k'}^f$ terms, which have the form of $\sin(\tau_{\max}x)/\sin(x)$ ($\rm sinc$-type behavior). The terms in \eqref{AnalyticalDist} are larger for the main lobe of the $\rm sinc$-type function, which makes $\tau_{\max}$ another factor that impacts the received distortion. If $\tau_{\max}$ is large, then the distortion power is dominated by the constant term $L^2/L^3$ in \eqref{AnalyticalDist} since the main lobe is narrow. Consequently, as $L$ increases, the received distortion power reduces. On the other hand, if $\tau_{\max}$ is small (the main lobe is wide), then the in-band distortion power, where $|k-k'|$ is small, is roughly proportional to $L^4/L^3$, which corresponds to the case that the received distortion gets stronger.

To sum up, the analytical framework provides insights into the received distortion power in massive MIMO systems in the presence of frequency-selective channels. The derived distortion power expression introduces the dependency of the distortion power to the number of MPCs and the maximum delay spread, which can be taken into account in communication system design. Firstly, the received SDR does not depend on the number of antennas. Secondly, as the number of MPCs increases, the beamforming gain of the nonlinear distortion is inversely proportional to the number of MPCs. Finally, it is also observed that as the maximum delay spread decreases, subcarrier channels decorrelate slowly, which yields different tendencies for in-band and OOB distortion powers.

\begin{figure*}[!t]
\centering
    \includegraphics[scale=0.5]{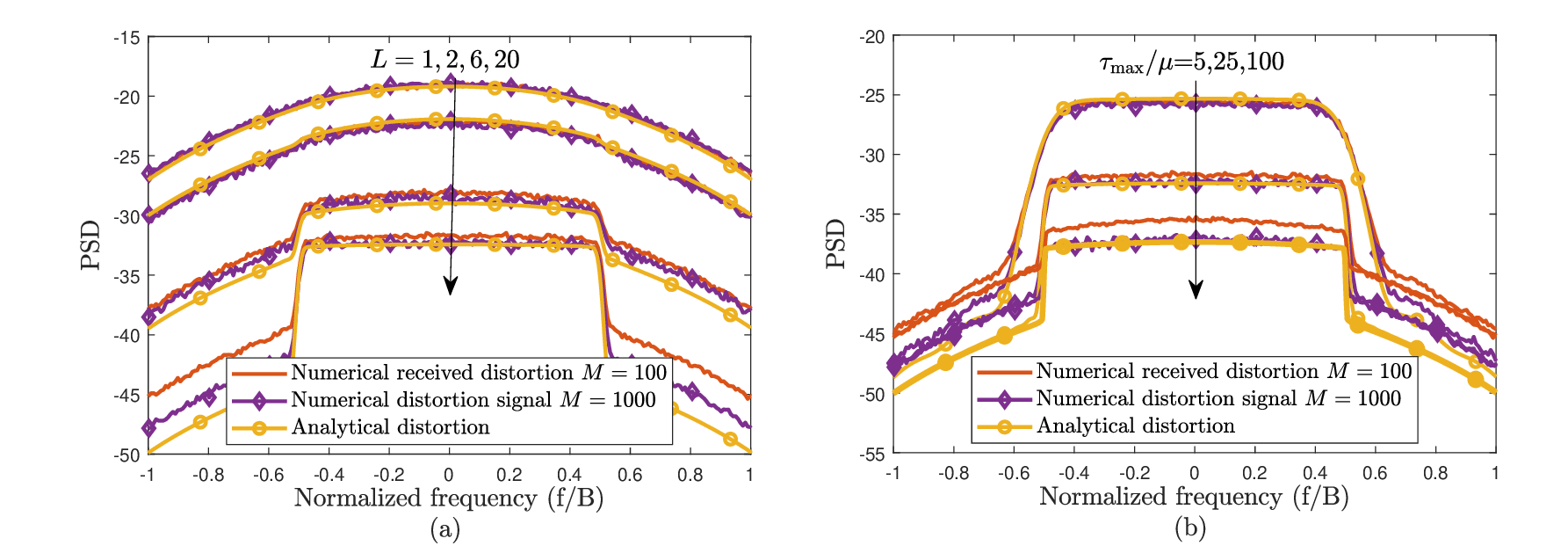}
    \caption{PSD of the received signal for different (a) number of MPCs for $\tau_{\max}/\mu =25$ and (b) delay spread for $L=20$.}
    \label{fig:PSD}
\end{figure*}

\section{Numerical Evaluations}

In this section, the derived distortion power expression in \eqref{AnalyticalDist} is verified, and the effects of frequency selectivity on both in-band and OOB distortion at the user location are investigated for different scenarios. To verify the derived distortion power expression, its proximity to the distortion power obtained by the Monte Carlo (MC) trials, which simulate a link-level communication system, is evaluated. In the simulations, a massive MIMO system with either $M=100$ or $M=1000$ antennas is investigated, and OFDM symbols having $N_s=1024$ subcarriers with oversampling factor $\mu=4$ is used to generate the time domain signals. The Rapp model is used to model the memoryless nonlinearity \cite{9804228}. For the given input $x[n]=|x[n]| e^{j\phi[n]}$, the output of the model is described as $\tilde{x}[n]= A(x[n]) e^{j\phi[n]}$, where

\begin{equation}
    A(x[n]) = \nu |x[n]| \left( 1 + \left(\frac{\nu |x[n]|}{r_o} \right)^{2\mu} \right)^{\frac{1}{2\mu}}. \label{RappModel}
\end{equation}
The Hermite polynomial coefficients in \eqref{RappModel} is extracted via linear regression to represent the nonlinearity as in \eqref{HermiteExpansion}.

\begin{figure*}[!t]
\centering
    \includegraphics[scale=0.5]{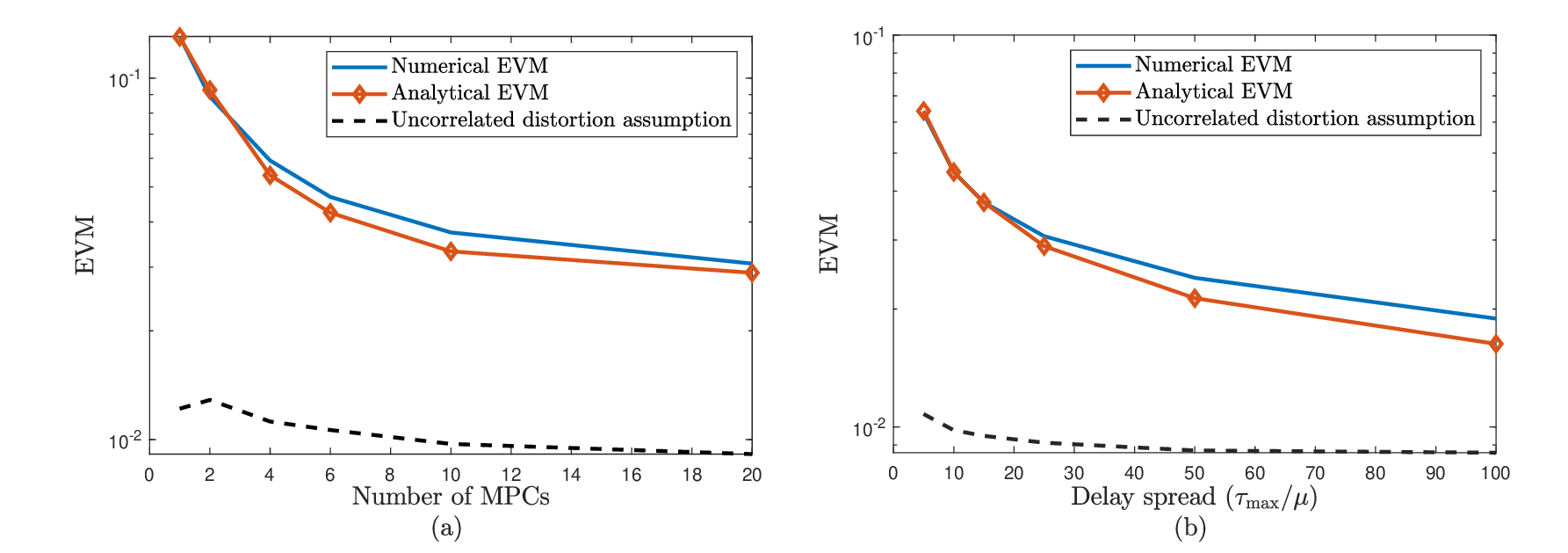}
    \caption{EVM performances for different (a) number of MPCs for $\tau_{\max}/\mu =25$ (b) delay spread for $L=20$.}
    \label{fig:EVM}
\end{figure*}

In Fig.~\ref{fig:PSD}, normalized PSDs of the received signals, for which the received signal power is scaled to be $1$, are shown to examine the effects of both the number of MPCs and the delay spread. It can be observed from Fig.~\ref{fig:PSD}(a) that the received distortion power tends to decrease as the number of MPCs increases; however, different behaviors are observed for the in-band and OOB distortion characteristics. The decay rate of the OOB distortion power is higher compared to that of in-band distortion, which is also verified by the analytical expression in \eqref{AnalyticalDist} since \eqref{AnalyticalDist} can indicate the distinctive characteristics for the in-band and OOB distortion as can be seen from In Fig.~\ref{fig:PSD}, and the obtained analytical distortion curves overlap with the numerical ones. The difference between the decay characteristics stems from the variation of the correlation between the subcarriers contained in $\epsilon_{k,k'}^{f2}$ and $\epsilon_{k,k',k''}^{ff}$ terms, which are described by the \emph{sinc}-type behavior. The reason is that terms in the summation \eqref{AnalyticalDist} are only non-zero for larger $k-k'$, $k-k''$ terms, which correspond to side-lobes of the \emph{sinc}-type function. However, in the case of in-band distortion, the terms are also non-negative for the neighborhood of $k$, which corresponds to the main lobe of the \emph{sinc} function. Eventually, the received distortion power is larger for the in-band distortion than for the OOB distortion. As the number of MPCs increases, the significance of the \emph{sinc}-related term becomes dominant over the constant term ($L$ or $L^2$). This yields a substantial difference between the in-band and OOB distortion powers. Fig.~\ref{fig:PSD}(a) also shows the PSD of the received signal for $M=1000$ to examine the nearly asymptotic case. It can be seen that as $M$ increases, the numerically  PSD converges to the analytical PSD for higher MPC numbers since the approximations in \eqref{Sd} are tighter. Therefore, it can be concluded that the obtained analytical received distortion expression provides an accurate asymptotic deterministic approximation. In addition, the analytical received distortion expression is a good approximation for many practical scenarios. It provides close approximations in practical parameter ranges, which can be observed from error vector magnitude (EVM) performance results in Fig.~\ref{fig:EVM}.

The impact of the maximum delay spread is demonstrated in Fig.~\ref{fig:PSD}(b). It can be observed that as the delay spread increases, the in-band distortion power converges to the OOB distortion power since the main lobe of the \emph{sinc}-type function gets narrower. A narrower main lobe reduces the number of significant terms in the summation \eqref{AnalyticalDist}, which yields a reduction in the received distortion power.

The effects of nonlinear distortion on the received signal quality are assessed via the EVM metric by using 64-QAM symbols for the case of $M=100$. We compute the theoretical EVM metric as the distortion-over-signal power ratio $1/{\rm{SDR}}[k]$. In Fig.~\ref{fig:EVM}, the impacts of the number of MPCs and delay spread on the EVM are shown. Besides, a \emph{hypothetical} case for the isotropic distortion radiation is included to emphasize the significance of the coherent beamforming of the distortion. From both Fig.~\ref{fig:PSD}(a) and Fig.~\ref{fig:PSD}(b), it is clear that neglecting coherent distortion beamforming yields a significant mismatch between the approximated and the exact EVM values. Therefore, a comprehensive received distortion power expression as in \eqref{AnalyticalDist} is required to evaluate the system performance accurately. It can be observed from Fig.~\ref{fig:EVM}(a) that the EVM performance improves with an increasing number of MPCs, which is consistent with what we observed for the PSD of the distortion. Lastly, from Fig.~\ref{fig:EVM}(b), it is also concluded that the EVM improves as the delay spread enlarges, which is compliant with the in-band distortion tendency shown in Fig.~\ref{fig:PSD}(b). %, and the theoretical EVM metric provides a close approximation for the numerically computed EVM.

\section{Conclusion}
This paper constructed a theoretical framework to analyze the impact of nonlinear distortion on downlink massive MIMO systems where the channel exhibits frequency-selective fading. A fully analytical received distortion power expression was derived and verified via Monte Carlo simulations. It is observed that as the number of MPCs increases, the level of nonlinear distortion that is coherently beamformed towards the intended receiver decreases. Hence, the total received distortion power reduces, which dilutes the effects of nonlinear distortion. Through the analysis, different tendencies are observed for the in-band and OOB distortion depending on how the correlation between subcarriers varies with the channel characteristics and number of antennas. Developed framework can be extended to multi-users scenarios and it can be employed to design system parameters to enhance spectral efficiency.

\section*{Acknowledgment}

M. B. Salman was supported by Scientific and Technical Research Council of Türkiye during his visit to the KTH Royal Institute of Technology, Stockholm, Sweden.
E. Bj\"ornson was supported by the Knut and Alice Wallenberg Foundation.

\bibliography{references} 
\bibliographystyle{IEEEtran}

\end{document}